\documentclass[twocolumn,prd,superscriptaddress,showpacs,floatfix,%
preprintnumbers,nofootinbib]{revtex4}

\usepackage{epsfig}
\usepackage{bm}

\newcommand{\I}{{\rm i}}

\newcommand{\D}{{\rm d}}

\begin{document}

\title{Decoherence
in supernova neutrino transformations
suppressed by deleptonization}

\author{Andreu Esteban-Pretel}
\affiliation{AHEP Group, Institut de F\'\i sica
 Corpuscular, CSIC/Universitat de Val\`encia,
 Edifici Instituts d'Investigaci\'o, Apt.\ 22085,
 46071 Val\`encia, Spain}

\author{Sergio Pastor}
\affiliation{AHEP Group, Institut de F\'\i sica
 Corpuscular, CSIC/Universitat de Val\`encia,
 Edifici Instituts d'Investigaci\'o, Apt.\ 22085,
 46071 Val\`encia, Spain}

\author{Ricard Tom\`as}
\affiliation{AHEP Group, Institut de F\'\i sica
 Corpuscular, CSIC/Universitat de Val\`encia,
 Edifici Instituts d'Investigaci\'o, Apt.\ 22085,
 46071 Val\`encia, Spain}

\author{Georg G.~Raffelt}
\affiliation{Max-Planck-Institut f\"ur Physik
(Werner-Heisenberg-Institut), F\"ohringer Ring 6, 80805 M\"unchen,
Germany}

\author{G{\"u}nter Sigl}
\affiliation{II.\ Institut f\"ur theoretische Physik, Universit\"at Hamburg\\
Luruper Chaussee 149, D-22761 Hamburg, Germany}
\affiliation{APC~\footnote{UMR 7164 (CNRS, Universit\'e Paris 7,
CEA, Observatoire de Paris)} (AstroParticules et Cosmologie),
10, rue Alice Domon et L\'eonie Duquet, 75205 Paris Cedex 13, France}

\date{28 December 2007}

\preprint{MPP-2007-60, IFIC/07-26}

\begin{abstract}
In the dense-neutrino region at 50--400~km above the neutrino sphere
in a supernova, neutrino-neutrino interactions cause large flavor
transformations. We study when the multi-angle nature of the
neutrino trajectories leads to flavor decoherence between different
angular modes. We consider a two-flavor mixing scenario between
$\nu_e$ and another flavor $\nu_x$ and assume the usual hierarchy
$F_{\nu_e}>F_{\bar\nu_e}>F_{\nu_x}=F_{\bar\nu_x}$ for the number
fluxes. We define
$\epsilon=(F_{\nu_e}-F_{\bar\nu_e})/(F_{\bar\nu_e}-F_{\bar\nu_x})$
as a measure for the deleptonization flux which is the one crucial
parameter. The transition between the quasi single-angle behavior
and multi-angle decoherence is abrupt as a function of $\epsilon$.
For typical choices of other parameters, multi-angle decoherence is
suppressed for $\epsilon\agt0.3$, but a much smaller asymmetry
suffices if the neutrino mass hierarchy is normal and the mixing
angle small. The critical $\epsilon$ depends logarithmically on the
neutrino luminosity. In a realistic supernova scenario, the
deleptonization flux is probably enough to suppress multi-angle
decoherence.
\end{abstract}


\maketitle

\section{Introduction}                        \label{sec:introduction}

In the dense neutrino flux emerging from a supernova (SN) core,
neutrino-neutrino refraction causes nonlinear flavor oscillation
phenomena that are unlike anything produced by ordinary
matter~\cite{Pastor:2002we, Sawyer:2005jk, Fuller:2005ae, Duan:2005cp,
Duan:2006an, Duan:2006jv, Hannestad:2006nj, Raffelt:2007yz,
Duan:2007mv, Raffelt:2007cb, Mirizzi2007}. The crucial phenomenon is a
collective mode of pair transformations of the form
$\nu_e\bar\nu_e\to\nu_x\bar\nu_x$ where $x$ represents some suitable
superposition of $\nu_\mu$ and $\nu_\tau$. This pair-wise form of
flavor transformation leaves the net flavor-lepton number flux
unchanged. Even an extremely small mixing angle is enough to trigger
this effect that is insensitive to the presence of ordinary matter
unless there is a Mikheyev-Smirnov-Wolfenstein (MSW) resonance in the
dense-neutrino region.

Collective pair transformations require a large neutrino density
{\it and\/} a pair excess of a given flavor. In typical SN models
one finds a hierarchy of number fluxes
$F_{\nu_e}>F_{\bar\nu_e}>F_{\nu_x}=F_{\bar\nu_x}$. The first part of
the hierarchy is caused by the deleptonization of the collapsed core
whereas the second is caused by the absence of charged-current
interactions for neutrino species other than $\nu_e$ and
$\bar\nu_e$. The neutrino fluxes streaming from a collapsed star
thus provide a natural environment for a flavor pair excess. On the
other hand, the SN core itself is characterized by a large $\nu_e$
chemical potential that enhances the $\nu_e$ density and suppresses
that of $\bar\nu_e$ so that here pair transformations cannot occur.
Likewise, the deleptonization burst immediately after core bounce
has an excess of $\nu_e$ and a depletion of $\bar\nu_e$
\cite{Kachelriess:2004ds}, suggesting that it is unaffected by
collective pair transformations.

\begin{figure}[b]
\includegraphics[angle=0,width=0.98\columnwidth]{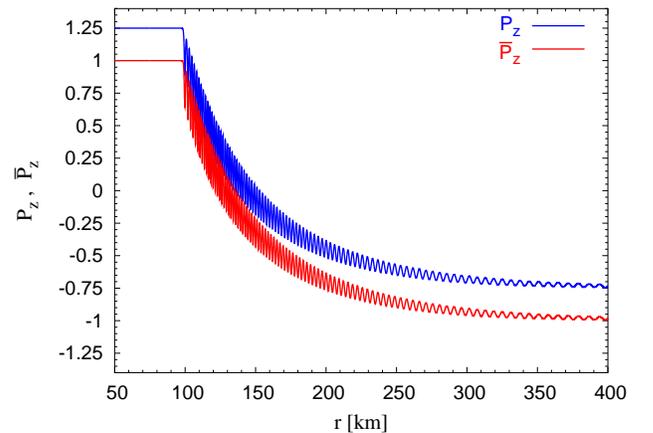}
\caption{Schematic evolution of the $z$-components of the total
polarization vectors for neutrinos and antineutrinos in a SN caused
by neutrino-neutrino interactions for the inverted-hierarchy example
described in the text.\label{fig:firstexample}}
\end{figure}

We illustrate collective pair conversions with a simple example in
Fig.~\ref{fig:firstexample}, assuming a typical SN neutrino luminosity
to be quantified later. We show the evolution of the $z$-components of
the global flavor polarization vector ${\bf P}$ for neutrinos and
$\bar{\bf P}$ for antineutrinos, where initially $P=|{\bf
  P}|=1+\epsilon$ with $\epsilon=0.25$ and $\bar P=|\bar{\bf P}|=1$.
We have assumed a monochromatic spectrum, that all neutrinos are
emitted at $45^\circ$ relative to the radial direction, the
atmospheric $\Delta m^2$, a small vacuum mixing angle
$\sin2\theta=10^{-3}$ to mimic the effect of ordinary matter.  and an
inverted mass hierarchy. For the normal hierarchy, no visible
evolution takes place.

Flavor oscillations do not change those parts of the flavor fluxes
that are already equal, only the transformation of the excess
$\bar\nu_e$ flux over the $\bar\nu_x$ flux is observable, and likewise
for neutrinos. The polarization vectors only represent this
excess. Therefore, without loss of generality we may set
$F_{\nu_x}=F_{\bar\nu_x}=0$ in our examples or equivalently, we may
picture $F_{\nu_e}$ and $F_{\bar\nu_e}$ to represent
$F_{\bar\nu_e}-F_{\bar\nu_x}$ and likewise for neutrinos.  Our chosen
parameters mean that at the neutrino sphere ($R=10$~km) the excess of
the $\nu_e$ flux over the $\nu_x$ flux is 25\% larger than the excess
of the $\bar\nu_e$ flux over the $\bar\nu_x$ flux ($\epsilon=0.25$).
$\bar P_z=+1$ then represents a pure $\bar\nu_e$ excess flux, $\bar
P_z=0$ represents equal excess fluxes of both flavors, and $\bar P_z=-1$ a pure
$\bar\nu_x$ excess flux, and analogous for $\nu_e$ with
$1\to1+\epsilon$.

The main features of Fig.~\ref{fig:firstexample} are nicely
explained after recognizing that the equations of motion can be
brought into a form where they are equivalent to a gyroscopic
pendulum~\cite{Hannestad:2006nj, Duan:2007mv}. The initial ``plateau
phase'' corresponds to synchronized oscillations or, in the pendulum
language, to a fast precession. We call the radius where this phase
ends the synchronization radius $r_{\rm synch}$. The decline with
``wiggles'' represents a nutation mode. The overall decline is
caused by the dilution of the neutrino flux and their increasing
collinearity with distance, corresponding to a decline of their
effective interaction energy.

One salient feature of Fig.~\ref{fig:firstexample} is that the
$\bar\nu_e$ flux completely converts to $\bar\nu_x$, whereas the
$\nu_e$ flux converts to $\nu_x$ only to the extent allowed by the
conservation of $P_z-\bar P_z=\epsilon$. This conservation is exact
in the mass basis that approximately coincides with the interaction
basis if the mixing angle is small.  In other words, only
$\nu_e\bar\nu_e$ pairs convert to $\nu_x\bar\nu_x$ pairs,
whereas the unpaired $\nu_e$ excess remains in its original
flavor~\cite{Hannestad:2006nj}.

The current-current nature of the weak interaction causes the
interaction energy to depend on $(1-\cos\theta)$ for two
trajectories with relative angle $\theta$. Therefore, neutrinos
emitted in different directions from a SN core experience different
refractive effects~\cite{Duan:2006an, Duan:2006jv}. As a result, one
would expect that their flavor content evolves differently, leading
to kinematical decoherence between different angular 
modes~\cite{Sawyer:2005jk}. Two of 
us have recently shown that this multi-angle decoherence is indeed
unavoidable in a ``symmetric gas'' of equal densities of neutrinos
and antineutrinos~\cite{Raffelt:2007yz}. Moreover, this effect is
self-accelerating in that an infinitesimal anisotropy is enough to
trigger an exponential run-away towards flavor equipartition, both
for the normal and inverted hierarchies.

In the SN context, however, it has been numerically observed that
the evolution is much more similar to the single-angle (or the
isotropic) case~\cite{Duan:2006an, Duan:2006jv}. The flux emitted by
a SN is extremely anisotropic. If one assumes $\nu\bar\nu$ symmetry,
flavor decoherence is swift and unavoidable. Therefore, the observed
suppression of multi-angle decoherence must be related to the
$\nu_e\bar\nu_e$ asymmetry that is generated by SN core
deleptonization.

\begin{figure*}
\includegraphics[width=0.85\textwidth]{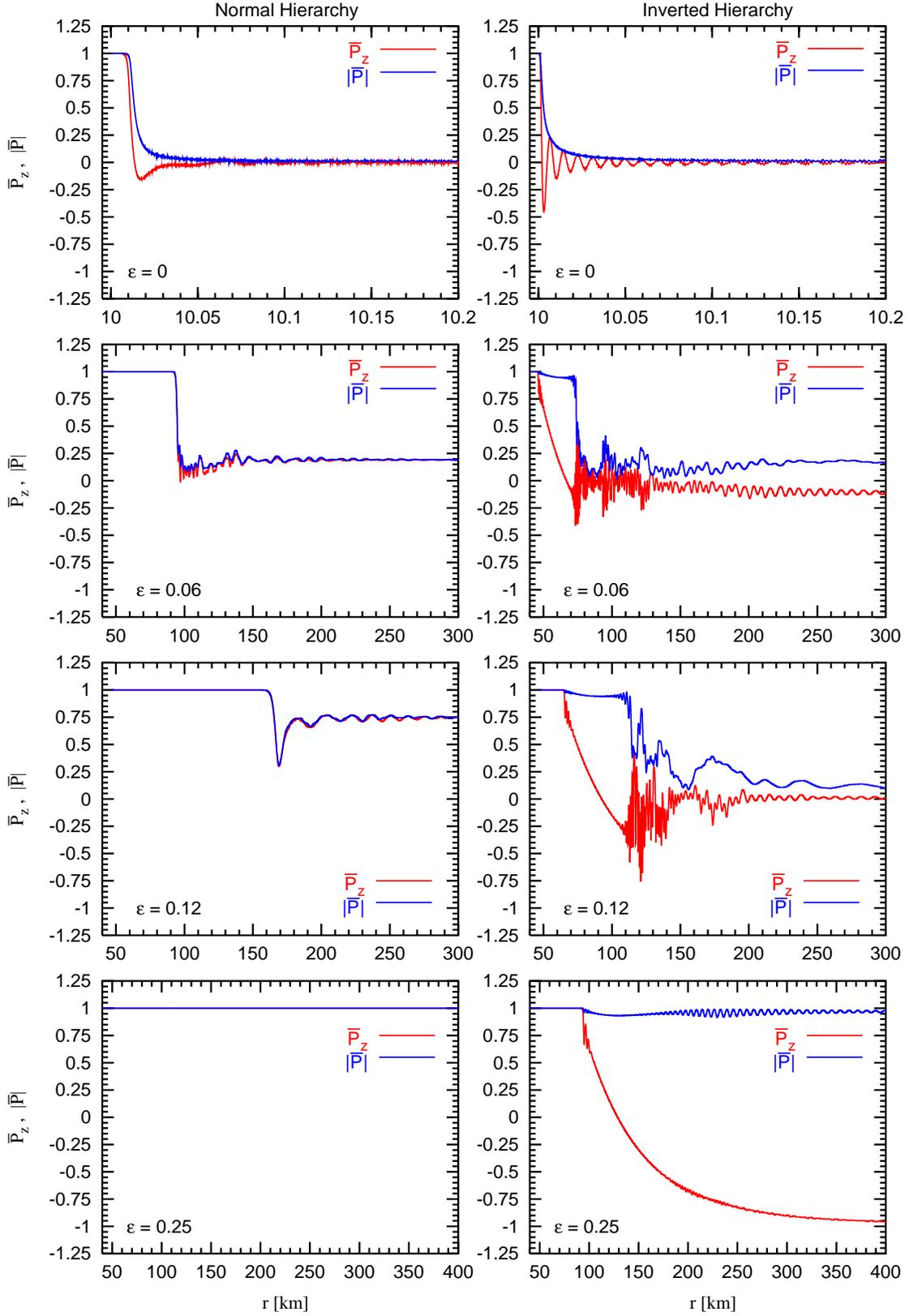}
\caption{Radial evolution of $\bar P_z$ in a schematic SN model as in
  Fig.~\ref{fig:firstexample}, but now for multi-angle neutrino
  emission at the neutrino sphere ($R=10$~km). In addition we show the
  length $\bar P=|\bar {\bf P}|$ as a measure of kinematical
  coherence. Left: normal hierarchy. Right: inverted hierarchy.  From
  top to bottom: $\epsilon=0$, 0.06, 0.12 and~0.25, where $\epsilon$
  is defined in Eq.~(\ref{eq:epsdefine}).\label{fig:example2}}
\end{figure*}

To illustrate this point we show in Fig.~\ref{fig:example2} a few
examples along the lines of Fig.~\ref{fig:firstexample}, but now for
multi-angle emission from the neutrino sphere that is again taken at
10~km. We consider different values of an asymmetry parameter that
we define as
\begin{equation}\label{eq:epsdefine}
 \epsilon=
 \frac{F(\nu_e)-F(\nu_x)}{F(\bar\nu_e)-F(\bar\nu_x)}-1
 =\frac{F(\nu_e)-F(\bar\nu_e)}{F(\bar\nu_e)-F(\bar\nu_x)}\,,
\end{equation}
where we have used $F(\nu_x)=F(\bar\nu_x)$. As mentioned earlier,
we can assume $F(\nu_x)=F(\bar\nu_x)=0$ at the neutrino sphere without
loss of generality. The left panels are for the normal hierarchy, the
right panels for the inverted hierarchy.

$P_z(r)-\bar P_z(r)=\epsilon$ is constant so that it is sufficient
to show $\bar P_z(r)$ alone. However, the length $\bar P=|\bar {\bf
P}|$ is no longer preserved: Complete kinematical decoherence among
the angular modes would cause $\bar P=0$. On the other hand, if
$\bar P=1$ remains fixed, this signifies that all modes evolve
coherently with each other. We use $\bar{\bf P}$ rather than ${\bf
P}$ because the former measures what happens to the $\nu_e\bar\nu_e$
pairs, whereas the latter also includes the conserved
$\nu_e$~excess.

In the top row we use $\epsilon=0$ (symmetric case). The flavor
content decoheres quickly as expected. Both the length and the
$z$-components of ${\bf P}$ and $\bar{\bf P}$ shrink to zero within
about 20~meters of the nominal neutrino sphere.

On the other extreme, we show in the bottom row the same for
$\epsilon=0.25$. In the normal hierarchy, nothing visible happens,
in analogy to the single-angle case.  In the inverted hierarchy, the
transformation is similar, but not identical, to the single-angle
case. The nutations wash out quickly. Shortly after exiting from the
synchronization phase, the length $\bar P$ shrinks a bit, but stays
almost constant thereafter. Clearly, some sort of multi-angle effect
has happened as we will discuss further in
Sec.~\ref{sec:decoherence}, but multi-angle decoherence has
certainly not occurred.

In the two middle rows we show intermediate cases with
$\epsilon=0.06$ and 0.12, respectively. For the inverted hierarchy,
these examples are qualitatively equivalent. The evolution is at first
similar to the single-angle case and analogous to $\epsilon=0.25$.
The nutations are washed out and the length $\bar P$ shrinks a
little bit after the synchronization radius. At some larger radius,
however, something new happens in that $\bar P$ suddenly shrinks
significantly, although not to zero, and there is a distinct feature
in the evolution of the $z$-component. Now we obtain partial
decoherence. The final flavor content is very different from the
single-angle case.

In the normal hierarchy, and for $\epsilon=0.06$, we obtain large
decoherence that begins abruptly at some radius far beyond $r_{\rm
synch}$. For the larger asymmetry $\epsilon=0.12$, the length $\bar
P$ also shrinks, but closely tracks $\bar P_z$. As we will see, this
case is somewhat like Phase~II of the inverted-hierarchy case, i.e.,
a certain amount of shrinking of the length of $\bar P$ and thus a
clear multi-angle effect, but no real decoherence.

Depending on the deleptonization flux, here represented by the
asymmetry parameter $\epsilon$, the system behaves very differently.
In particular, for the inverted hierarchy it is striking that there
are either two or three distinct phases. We always have the initial
synchronized phase at large neutrino densities. Next, there is
always the quasi single-angle pair-transformation phase at distances
larger than $r_{\rm synch}$. Just beyond this radius, the global
polarization vectors quickly shrink by a small amount, but then
stabilize immediately. Finally, if $\epsilon$ is below some critical
value, there is a sharp transition to a third phase where the
different angular modes decohere significantly, but not completely.
The practical outcome for the flavor fluxes emerging from the
dense-neutrino region is very different depending on $\epsilon$. The
transition between these regimes is abrupt, a small change of
$\epsilon$ is enough to cause one or the other form of behavior.

While these phenomena call for an analytic quantitative
understanding, we are here less ambitious, but more practical. We
study numerically for which range of parameters the different forms
of behavior occur. Towards this goal we first set up, in
Sec.~\ref{sec:setup}, our conventions, the equations of motion for a
spherically symmetric system, and establish the connection between
the parameters of our schematic model with those of a realistic SN
scenario. In Sec.~\ref{sec:decoherence} we describe in more detail
what happens in the different phases of evolution diagnosed in
Fig.~\ref{fig:example2} and identify useful measures of decoherence.
In Sec.~\ref{sec:parameters} we investigate the role of our various
model parameters in determining if the system kinematically
decoheres. We discuss our findings and conclude in
Sec.~\ref{sec:conclusions}. In Appendix~\ref{app:eom} we derive the
equations of motion adapted to spherical symmetry.

\section{Setup of the problem}                       \label{sec:setup}

\subsection{Equations of motion}

To study the flavor evolution of the neutrino flux emitted by a SN
core we solve numerically the equations of motion for the
flavor-dependent number fluxes, assuming spherical symmetry. We
always work in a two-flavor scenario between $\nu_e$ and another
flavor $\nu_x$, characterized by the atmospheric $\Delta m^2$ and by
a vacuum mixing angle $\theta$ that is taken to represent the
unknown 13-mixing angle.

Our fundamental quantities are the flux matrices in flavor space
${\sf J}_r$ that depend on the radial coordinate $r$
(Appendix~\ref{app:eom}). The diagonal entries represent the total
neutrino number fluxes through a sphere of radius~$r$. In the
absence of oscillations, ${\sf J}_r$ would not depend on the radius at
all. The flux matrices are represented by polarization vectors ${\bf
P}_r$ in the usual way,
\begin{eqnarray}\label{eq:pol}
 {\sf J}_r&=&
 \frac{F(\nu_e)+F(\nu_x)}{2}
 +\frac{F(\bar\nu_e)-F(\bar\nu_x)}{2}\,
 {\bf P}_r\cdot\bm{\sigma}\,,
 \nonumber\\
 \bar{\sf J}_r&=&
 \frac{F(\bar\nu_e)+F(\bar\nu_x)}{2}
 +\frac{F(\bar\nu_e)-F(\bar\nu_x)}{2}\,
 \bar{\bf P}_r\cdot\bm{\sigma}\,,
\end{eqnarray}
where $\bm\sigma$ is the vector of Pauli matrices. Antineutrino
quantities are always denoted with an overbar. The number fluxes
$F(\nu)$ are understood at the neutrino sphere. In both equations
the term proportional to the polarization vector is normalized to
the antineutrino flux. As a consequence, at the neutrino sphere we
have the normalization
\begin{equation}
 P=|{\bf P}|=1+\epsilon
 \hbox{\quad and\quad}
 \bar P=|\bar{\bf P}|=1\,.
\end{equation}
In this way, we treat the excess flux from deleptonization as an
adjustable parameter without affecting the baseline flux of
antineutrinos.

The diagonal part of the flux matrices is conserved and irrelevant
for flavor oscillations. The polarization vector ${\bf P}_r$ only
captures the {\it difference\/} between the flavor fluxes. For this
reason we have defined the asymmetry $\epsilon$ in terms of the flux
differences.

Multi-angle effects are at the focus of our study. We label different
angular modes with
\begin{equation}
u=\sin^2\vartheta_R\,,
\end{equation}
where $\vartheta_R$ is the zenith angle at the neutrino sphere $r=R$
of a given mode relative to the radial direction. The parameter
$u$ is fixed for every trajectory whereas the physical zenith angle
$\vartheta_r$ at distance $r$ varies. Therefore, using the local
zenith angle to label the modes would complicate the equations.

We will consider two generic angular distributions for the modes. In
the multi-angle case we assume that the neutrino radiation field is
``half isotropic'' directly above the neutrino sphere, i.e., all
outward moving modes are equally occupied as expected for blackbody
emission. This implies (Appendix~\ref{app:eom})
\begin{equation}
{\bf P}_{u,r}=\D {\bf P}_{r}/\D u=  {\rm const.}
\end{equation}
at $r=R$ for $0\leq u\leq 1$. Note that $u=0$ represents radial
modes, $u=1$ tangential ones. The other generic distribution is the
single-angle case where all neutrinos are taken to be launched at
$45^\circ$ at the neutrino sphere so that $u=1/2$ for all neutrinos.

For a monochromatic energy distribution, the equations of motion in
spherical symmetry are (Appendix~\ref{app:eom})
\begin{widetext}
\begin{eqnarray}\label{eq:eom5}
 \partial_r{\bf P}_{u,r}&=&
 +\frac{\omega {\bf B}\times{\bf P}_{u,r}}{v_{u,r}}
 +\frac{\lambda_r{\bf L}\times{\bf P}_{u,r}}{v_{u,r}}
 +\mu\,\frac{R^2}{r^2}
 \left[\left(\int_0^1\D u'\,
 \frac{{\bf P}_{u',r}-\bar{\bf P}_{u',r}}{v_{u',r}}\right)
 \times\left(\frac{{\bf P}_{u,r}}{v_{u,r}}\right)
 -({\bf P}_r-\bar{\bf P}_r)\times{\bf P}_{u,r}\right]\,,
 \nonumber\\*
 \partial_r\bar{\bf P}_{u,r}&=&
 -\frac{\omega {\bf B}\times\bar{\bf P}_{u,r}}{v_{u,r}}
 +\frac{\lambda_r{\bf L}\times\bar{\bf P}_{u,r}}{v_{u,r}}
 +\mu\,\frac{R^2}{r^2}
 \left[\left(\int_0^1\D u'\,
 \frac{{\bf P}_{u',r}-\bar{\bf P}_{u',r}}{v_{u',r}}\right)
 \times\left(\frac{\bar{\bf P}_{u,r}}{v_{u,r}}\right)
 -({\bf P}_r-\bar{\bf P}_r)\times\bar{\bf P}_{u,r}\right]\,,
\end{eqnarray}
\end{widetext}
where the radial velocity of mode $u$ at radius $r$ is
\begin{equation}
v_{u,r}=\sqrt{1-u\,R^2/r^2}\,.
\end{equation}
Further, $\omega=|\Delta m^2/2E|$ is the vacuum oscillation
frequency, taken to be positive. ${\bf
B}=(\sin2\theta,0,\pm\cos2\theta)$ where the mixing angle $\theta$
is usually taken to be small. $B_z<0$ corresponds to the normal
hierarchy, $B_z>0$ to the inverted hierarchy. ${\bf L}$ is a unit
vector in the $z$-direction because we work in the interaction
basis. The matter density is represented by
\begin{equation}
\lambda_r=\sqrt2 G_{\rm F}\,[n_{e^-}(r)-n_{e^+}(r)]\,.
\end{equation}
The strength of the neutrino-neutrino interaction is parameterized
by
\begin{equation}\label{eq:mudefine}
 \mu=\sqrt2 G_{\rm F}\left(F_{\bar\nu_e}^R-F_{\bar\nu_x}^R\right)\,,
\end{equation}
where the fluxes are taken at the neutrino sphere with radius~$R$.

The somewhat complicated structure of the equations arises from
projecting the evolution of each mode on the radial direction. This
is still very much simpler than following the evolution on every
trajectory as a function of distance (or time) on that trajectory.
We have here a closed set of differential equations that is not hard
to solve numerically.

We show in Appendix~\ref{app:eom} that for $r\gg R$,
the vacuum and matter oscillation terms take on the familiar
plane-wave form because at large distances all neutrinos essentially
move on radial trajectories. The neutrino-neutrino term falls off as
$r^{-4}$, in agreement with the previous literature.


\subsection{Schematic supernova model}                  \label{sec:sn}

We always consider a two-flavor oscillation scenario driven by the
atmospheric $\Delta m^2=1.9$--$3.0\times10^{-3}~{\rm eV}^2$.
Assuming $\langle E_\nu\rangle=15$~MeV, the
oscillation frequency is $\omega=\hbox{0.3--0.5}~{\rm km}^{-1}$. To
be specific, we use
\begin{equation}\label{eq:omegadefine}
\omega=\left\langle\frac{\Delta m^2}{2E}\right\rangle=
0.3~{\rm km}^{-1}
\end{equation}
as a benchmark value in the monochromatic model.

The total energy output of a SN is around $3\times10^{53}~{\rm
erg}$, corresponding to $0.5\times10^{53}~{\rm erg}$ in each of the
six neutrino species if we assume approximate equipartition of the
emitted energy. If this energy is emitted over 10~s, the average
luminosity per flavor would be $0.5\times10^{52}~{\rm erg/s}$.
However, at early times during the accretion phase, the luminosity
in the $\bar\nu_e$ flavor can exceed $3\times10^{53}~{\rm
erg/s}$~\cite{Raffelt:2003en}. As our baseline estimate we use
\begin{eqnarray}\label{eq:muestimate}
 \mu&=&7\times10^5~{\rm km}^{-1}\\
 &\times&
 \left(\frac{L_{\bar\nu_e}}{\langle E_{\bar\nu_e}\rangle}
 -\frac{L_{\bar\nu_x}}{\langle E_{\bar\nu_x}\rangle}\right)
 \frac{15~{\rm MeV}}{10^{52}~{\rm erg/s}}\,
 \left(\frac{10~{\rm km}}{R}\right)^2\,.\nonumber
\end{eqnarray}
This is significantly larger than the assumptions of previous
studies~\cite{Duan:2006an, Hannestad:2006nj}. Unless otherwise
stated, we always use the benchmark values for the different
parameters summarized in Table~\ref{tab:benchmark}.

\begin{table}
\caption{Default values for our model parameters.
\label{tab:benchmark}}
\begin{ruledtabular}
\begin{tabular}{lll}
Parameter&Standard value&Definition\\
\hline
$\epsilon$&0.25&Eq.~(\ref{eq:epsdefine})\\
$\mu$&$7\times10^5~{\rm km}^{-1}$&Eq.~(\ref{eq:mudefine})\\
$\omega$&$0.3~{\rm km}^{-1}$&Eq.~(\ref{eq:omegadefine})\\
$\sin2\theta$&$10^{-3}$&---\\
\end{tabular}
\end{ruledtabular}
\end{table}

In our calculations we always take the neutrino sphere at the radius
$R=10$~km. Of course, the physical neutrino sphere is not a
well-defined concept. Therefore, the radius $R$ simply represents
the location where we fix the inner boundary condition. However,
essentially nothing happens until the synchronization radius $r_{\rm
synch}\gg R$ because the in-medium mixing angle is extremely small
and both neutrinos and antineutrinos simply precess around~${\bf
B}$. Therefore, as far as the vacuum and matter oscillation terms
are concerned, it is almost irrelevant where we fix the inner
boundary condition.

Not so for the neutrino-neutrino term because we also fix the
angular distribution at $r=R$. While the $r^{-2}$ scaling
from flux dilution is unaffected by the radius for the inner
boundary condition, the ``collinearity suppression'' also scales as
$(R/r)^{2}$ for $r\gg R$. If we fix a half-isotropic distribution or
a single angle of $45^\circ$ at a larger radius $R'$, the new inner
boundary condition essentially amounts to
$\mu\to\mu'=\mu\,(R'/R)^2$. In the early phase after
bounce $R'=30$~km could be more realistic, leading to a $\mu$ value
almost an order of magnitude larger. Evidently, $\mu$ is a rather
uncertain model parameter that can differ by orders of magnitude
from our benchmark value.

However, collective pair conversions only begin at $r_{\rm synch}$
where $\mu$ is so small that synchronization ends. Therefore, the
main impact of a modified $\mu$ is to change $r_{\rm synch}$ and
thus to push the collective pair conversions to larger radii. The
oscillations are synchronized if~\cite{Hannestad:2006nj}
\begin{equation}
\frac{\mu}{\omega}>\frac{2}{(1-\sqrt{1+\epsilon})^2}\,.
\end{equation}
In our single-angle case we find from Eq.~(\ref{eq:muvariation})
that the effective neutrino-neutrino interaction strength varies at
large distances as
\begin{equation}
\mu_{\rm eff}(r)=\mu\,\frac{R^4}{2r^4}\,.
\end{equation}
Therefore, the synchronization radius is
\begin{eqnarray}
 \frac{r_{\rm synch}}{R}&=&
 \left(\frac{\sqrt{1+\epsilon}-1}{2}\right)^{1/2}
 \left(\frac{\mu}{\omega}\right)^{1/4}
 \nonumber\\*
 &\approx&\frac{\sqrt\epsilon}{2}\,
 \left(\frac{\mu}{\omega}\right)^{1/4}\,.
\end{eqnarray}
The second line assumes $\epsilon\ll 1$. If we use our benchmark
values $\omega=0.3~{\rm km}^{-1}$, $\mu=7\times10^5~{\rm km}^{-1}$,
$R=10$~km and $\epsilon=0.25$, we find $r_{\rm synch}=95$~km,
corresponding well, for example, to Fig.~\ref{fig:firstexample}. In
any event, if $\mu$ is taken to be uncertain by two orders of
magnitude, $r_{\rm synch}$ only changes by a factor of 3.

The total electron lepton number emitted from a collapsed SN core is
about $3\times10^{56}$. On the other hand, assuming that each
neutrino species carries away $0.5\times10^{53}~{\rm erg}$ with an
average energy of 15~MeV, the SN core emits about $2\times10^{57}$
neutrinos in each of the six species. In this simplified picture,
the SN emits on average about 15\% more $\nu_e$ than $\bar\nu_e$.
However, in the oscillation context we need the excess of
$F_{\nu_e}-F_{\nu_x}$ relative to the same quantity for
antineutrinos as defined in Eq.~(\ref{eq:epsdefine}). The true value
of $\epsilon$ thus depends sensitively on the detailed fluxes and
spectra of the emitted neutrinos. The asymmetry parameter is large
when the first hierarchy in
$F_{\nu_e}>F_{\bar\nu_e}>F_{\bar\nu_x}=F_{\nu_x}$ is large and/or
the second hierarchy is small. Even if $F_{\bar\nu_x}$ is as small
as half of $F_{\bar\nu_e}$, the asymmetry $\epsilon$ would be as
large as 30\%, even when $F_{\nu_e}$ exceeds $F_{\bar\nu_e}$ by only
15\%.

\subsection{Numerical multi-angle decoherence and the inner boundary
condition}   \label{sec:numerical}

One important and somewhat confusing complication of numerically
solving the equations of motion is the phenomenon of numerical
multi-angle decoherence. To integrate Eq.~(\ref{eq:eom5}) one needs
to work with a finite number of angular modes, equivalent to
coarse-graining the phase space of the system. If the number of
angular bins is chosen smaller than some critical number $N_{\rm
min}$, multi-angle decoherence occurs for $r<r_{\rm synch}$, where
physically it is not possible and does not occur for a fine-grained
calculation. This phenomenon is shown, for example, in Fig.~3 of
Ref.~\cite{Duan:2006an}. It is not caused by a lack of numerical
precision, but a result of the coarse-graining of phase space. A
related phenomenon is recurrence as discussed in the context of
multi-angle decoherence in Ref.~\cite{Raffelt:2007yz}.

In other words, a coarsely grained multi-angle system behaves
differently than a finely grained one. A smaller mixing angle
reduces $N_{\rm min}$, a larger neutrino-neutrino interaction
strength increases it. It should be possible to estimate $N_{\rm
min}$ from first principles, but for the moment we need to rely on
trial and error.

Starting the integration at $r=R$ is doubly punishing because the
fast oscillations of individual modes caused by a large $\mu$
requires many radial steps for the numerical integration and
avoiding numerical decoherence requires a large number of angular
modes. On the other hand, in this region nothing but fast
synchronized oscillations take place that have no physical effect if
the mixing angle is small. Using a larger radius as a starting point
for the integration avoids both problems and does not modify the
overall flavor evolution at larger distances.

From the physical perspective, the ``neutrino sphere'' is not a
well-defined concept because different energy modes and different
species decouple at different radii, and in any case, each
individual neutrino scatters last at a different radius. If the
exact inner boundary condition would matter, we would need to solve
the full kinetic equations, including neutral-current and
charged-current collisions. It is the beauty of the
neutrino-neutrino flavor transformation problem that the real action
begins at $r_{\rm synch}$, significantly outside the neutrino
sphere. Our approach of reducing the equations of motion to the
refractive terms is only self consistent because the exact location
of the inner boundary condition is irrelevant.

In summary, the nominal neutrino sphere at $R=10$~km is nothing but
a point of reference where we normalize the fluxes and fix the
angular distribution. As a starting point for integration we
typically use $r_0=0.75\,r_{\rm synch}$. A few hundred angular modes
are then usually enough to avoid numerical decoherence.

We note, however, that the normal-hierarchy cases are more sensitive
to both the number of angular bins and the starting radius for the
integration. It can happen that a case that looks like the
$\epsilon=0.12$ example in Fig.~\ref{fig:example2}, which shows a
mild shrinking of the polarization vector, can become ``more
coherent'' by choosing a smaller starting radius which then may also
require a larger number of modes. For the normal hierarchy, the
different multi-angle cases are less cleanly separated from each other
than in the inverted hierarchy in that the transition is less abrupt
as a function of $\epsilon$.

When physical multi-angle decoherence occurs (e.g.~the middle rows
of Fig.~\ref{fig:example2}), a much larger number of modes is needed
to provide reproducible results. However, we are here not interested
in the exact final outcome, we are mostly interested in the range of
parameters that lead to decoherence. Therefore, massive computation
power is not needed for our study.

For those cases where we include a non-trivial spectrum of energies
we also need energy bins. A distribution of energies does not lead
to kinematical decoherence in the context of collective neutrino
oscillations~\cite{Hannestad:2006nj} so that the number of energy
bins is not a crucial parameter. Of course, to resolve the
energy-dependent behavior and especially the spectral
splits~\cite{Duan:2006an, Duan:2007mv, Raffelt:2007cb, Mirizzi2007},
a sufficiently fine-grained binning is required. It provides better
resolution, but not a qualitatively different form of behavior.

\section{Coherent evolution vs.\ decoherence}
\label{sec:decoherence}

\begin{figure*}
\includegraphics[width=0.85\textwidth]{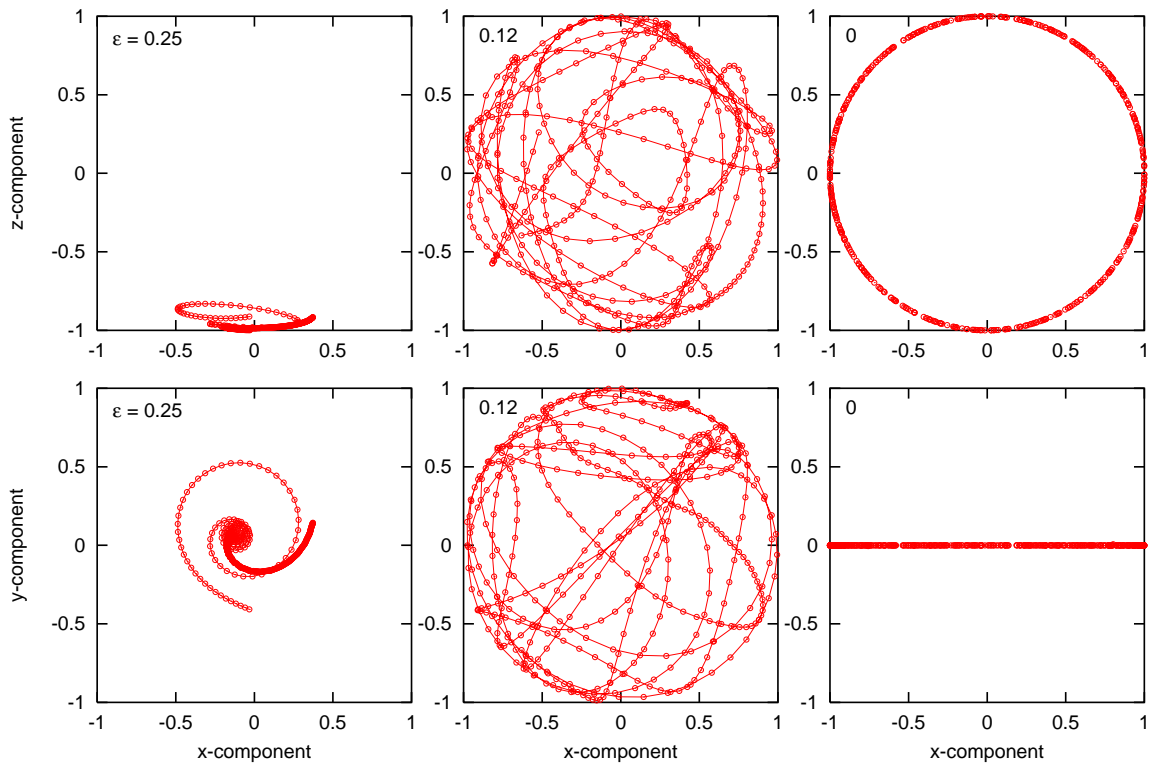}
\caption{Final location on the unit sphere of 500 antineutrino
  polarization vectors for our standard parameters and the inverted
  hierarchy. The top row is the ``side view'' ($x$-$z$-components),
  the bottom row the ``top view'' ($x$-$y$-components). Left:~quasi
  single-angle case ($\epsilon=0.25$). Middle: decoherent case
  ($\epsilon=0.12$). Right: symmetric system
  ($\epsilon=0$).\label{fig:footprints1}}
\end{figure*}

\begin{figure*}
\includegraphics[width=0.85\textwidth]{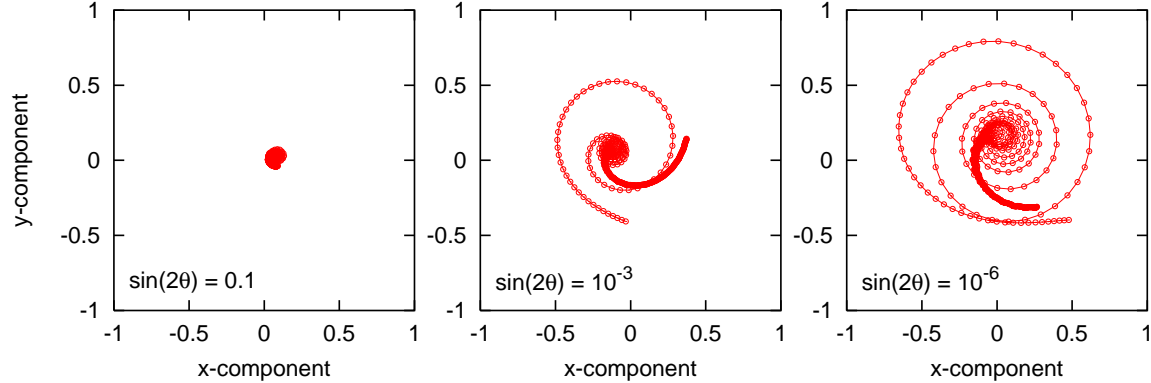}
\caption{Same as Fig.~\ref{fig:footprints1}, now only top views for
quasi single-angle cases with the mixing angles $\sin2\theta=0.1$,
$10^{-3}$ and $10^{-6}$ from left to right. The middle panel is
identical with the bottom left panel of
Fig.~\ref{fig:footprints1}.\label{fig:footprints4}}
\end{figure*}

\subsection{Different forms of evolution}

Before investigating the conditions for decoherence among angular
neutrino modes we first take a closer look at what happens in
the different cases shown in Fig.~\ref{fig:example2}. Considering
first the quasi single-angle case with the asymmetry
$\epsilon=0.25$, some insight is gained by looking at the final
state of the evolution at some large radius where the
neutrino-neutrino effects have completely died out and all modes
simply perform vacuum oscillations. In the left-hand panels of
Fig.~\ref{fig:footprints1} we show the end state of 500~polarization
vectors, representing modes uniformly spaced in the
angular coordinate $u$. In the upper
panel we show the final state in the $x$-$z$-plane (``side view''),
in the lower panel in the $x$-$y$-plane (``top view'').

\begin{figure}
\includegraphics[width=0.6\columnwidth]{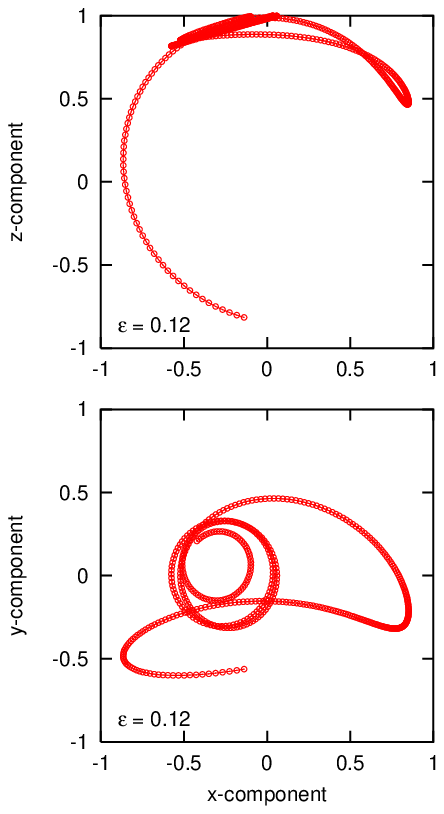}
\caption{Same as Fig.~\ref{fig:footprints1}, here for the normal
hierarchy and $\epsilon=0.12$.\label{fig:footprints2}}
\end{figure}

Initially, all polarization vectors are aligned in the flavor
direction. At the beginning of the pair transformation phase at
$r_{\rm synch}$, some are peeled off, forming a spiral structure
that is easily gleaned from the left panels of
Fig.~\ref{fig:footprints1}. This structure continues to evolve
almost as in the single-angle case, i.e., once established it moves
almost like a rigid body and eventually orients itself in the
negative ${\bf B}$-direction. Of course, it continues to rotate
around the ${\bf B}$ direction even at large radii because of vacuum
oscillations.

The spiral structure is different depending on the mixing angle. We
illustrate this in Fig.~\ref{fig:footprints4} where we show the top
view in analogy to the lower-left panel of Fig.~\ref{fig:footprints1}
for different choices of mixing angle. For a large $\sin2\theta$,
the polarization vectors stay close to each other. For a smaller
$\sin2\theta$, the spiral spreads over a larger solid angle and has
more windings. We recall that a smaller $\sin2\theta$ also has the
effect of causing a larger nutation depth of the flavor
pendulum~\cite{Raffelt:2007yz}.

Now turn to the quasi decoherent case with $\epsilon=0.12$.
Initially the same happens, but at the ``decoherence radius'' the
spiral structure dissolves almost instantaneously. The polarization
vectors enter a complicated structure as illustrated by the end
state (central panels of Fig.~\ref{fig:footprints1}). Moreover, they
are spread out all over the unit sphere, having both positive and
negative $z$-components. This structure looks different for
different choices of $\sin2\theta$ and $\epsilon$. However, once a
sufficient number of polarization vectors is used, it is
reproducible. For $\epsilon=0.06$ the picture would be qualitatively
similar.

Finally we show the fully symmetric case ($\epsilon=0$) in the
right-hand panels. Here decoherence is fast and complete. For a
small mixing angle, all polarization vectors are confined to the
$x$-$z$-plane. They distribute themselves on a circle in that plane.

For the normal hierarchy, we show in Fig.~\ref{fig:footprints2} as
an explicit example the $\epsilon=0.12$ case of
Fig.~\ref{fig:example2} that showed a clear multi-angle effect
without strong decoherence. Once more we find a spiral structure.
Most polarization vectors remain oriented roughly in their original
direction, but in this case also with a tail of a few polarization
vectors reversed. The quasi decoherent case ($\epsilon=0.06$) and
the symmetric system produce similar final pictures as the
corresponding cases of the inverted hierarchy.

\subsection{Measures of decoherence}

\begin{figure}[b]
\includegraphics[width=0.95\columnwidth]{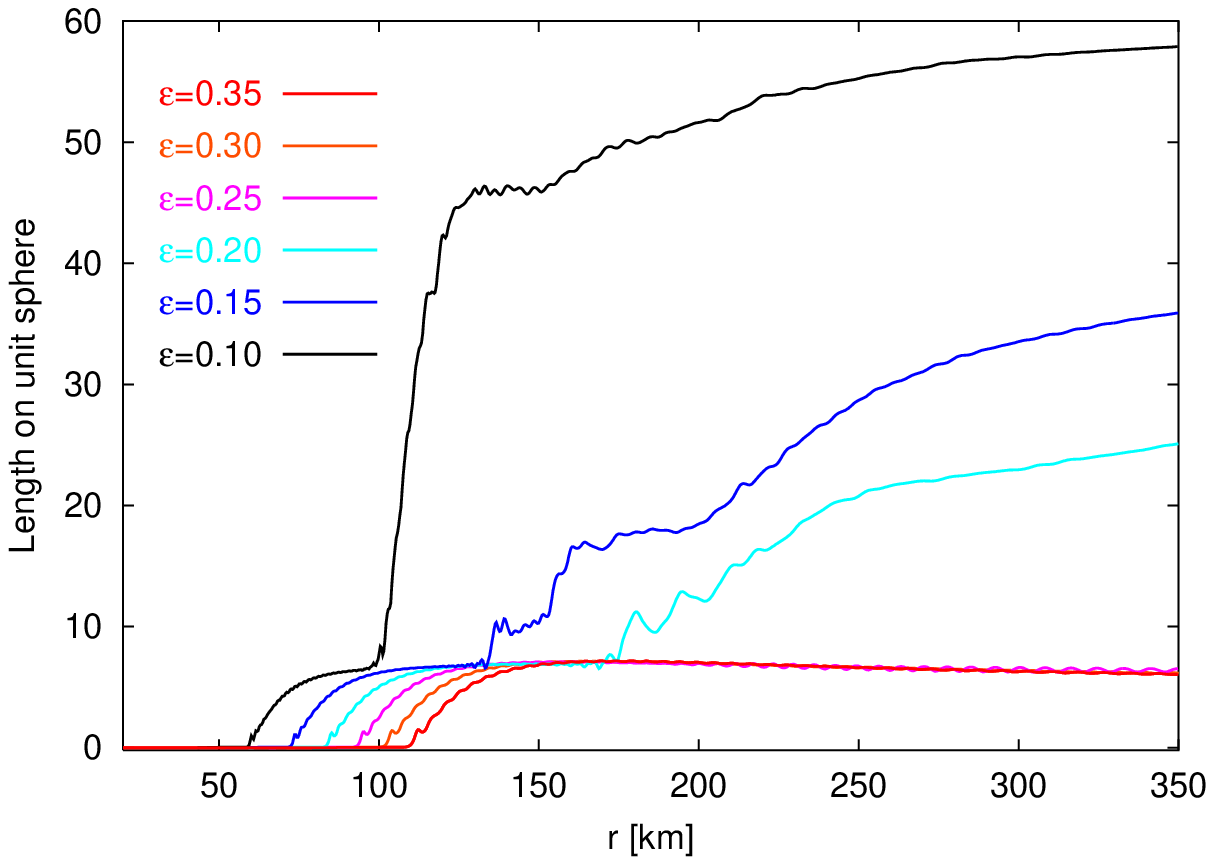}
\caption{Evolution of the length of the one-dimensional subspace
occupied by the polarization vectors for our standard inverted
hierarchy case, taking a series of different asymmetries
$\epsilon$. The length grows to larger values for smaller asymmetries.
\label{fig:phasespace}}
\end{figure}

Even in the quasi-decoherent cases the unit sphere is not uniformly
filled with polarization vectors. Rather, in the mono-energetic case
considered here, the occupied phase space is a one-dimensional
subspace of the unit sphere. It is parameterized by the angular
variable $u$ and shows a clear line-like structure. This picture
suggests to use the length of this line on the unit sphere as
another global measure besides the length $\bar P$ to discriminate
between different modes of evolution~\cite{Raffelt:2007yz}. In a
numerical run with discrete angular bins, this quantity is simply
the sum of the angles between neighboring polarization vectors. In
Fig.~\ref{fig:phasespace} we show this quantity for the indicated
values of $\epsilon$ as a function of radius for our
inverted-hierarchy examples.

At the radius $r_{\rm synch}$ where the spiral forms, the length on
the unit sphere quickly increases from~0 to a value that is almost
independent of $\epsilon$, but depends on the mixing angle. For
smaller $\sin2\theta$ it is larger, corresponding to the spiral
having more windings as indicated earlier. Later, this length stays
practically constant, reflecting that the spiral structure, once
established, does not change much except tilting toward the negative
${\bf B}$-direction and precessing around it.

When $\epsilon$ is smaller than a critical value, at the
``decoherence radius'' a sudden second growth phase shoots up from
the plateau of these curves. For smaller $\epsilon$, the final
length is longer, representing a more ``phase-space filling'' line
on the unit sphere.

Note, however, that for $\epsilon$ close to zero, the line does not
fill the unit sphere, but essentially stays in a narrow band. In the
perfectly symmetric case, the motion of all polarization vectors is
essentially confined to the $x$-$z$-plane, i.e., the polarization
vectors distribute themselves over a great circle on the sphere as
shown in the right panels of Fig.~\ref{fig:footprints1}.

\section{Role of model parameters} \label{sec:parameters}

\subsection{Ordinary matter} \label{sec:matter}

\begin{figure}[b]
\includegraphics[angle=0,width=0.95\columnwidth]{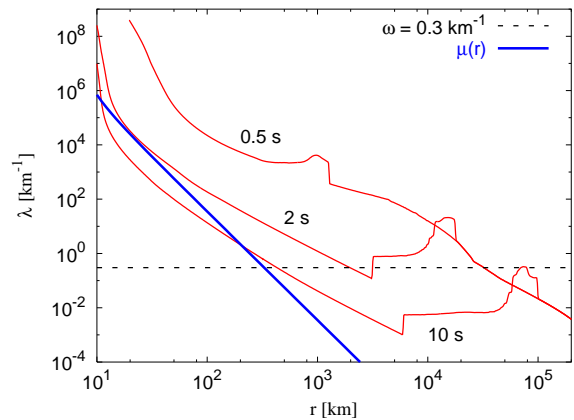}
\caption{Typical matter density profiles from numerical simulations of
the Garching group at the indicated times after core
bounce~\cite{Arcones:2006uq}. For comparison we also show our
benchmark value $\omega=0.3~{\rm km}^{-1}$ and $\mu(r)$ for a typical
case.\label{fig:profiles}}
\end{figure}

We now explore how various model parameters influence the behavior of
the system. In the examples so far we have ignored matter because its
effect is mainly to suppress the vacuum mixing angle. We here make
this argument more precise. In Fig.~\ref{fig:profiles} we show typical
matter density profiles, expressed in terms of the matter oscillation
frequency $\lambda(r)$, from numerical simulations of the Garching
group for different times after collapse~\cite{Arcones:2006uq}. For
comparison we also show $\mu(r)$ with $\mu(R)=7\times10^5~{\rm
km}^{-1}$ and a radial variation in analogy to
Eq.~(\ref{eq:muvariation}).

\begin{figure}[b]
\includegraphics[angle=0,width=0.95\columnwidth]{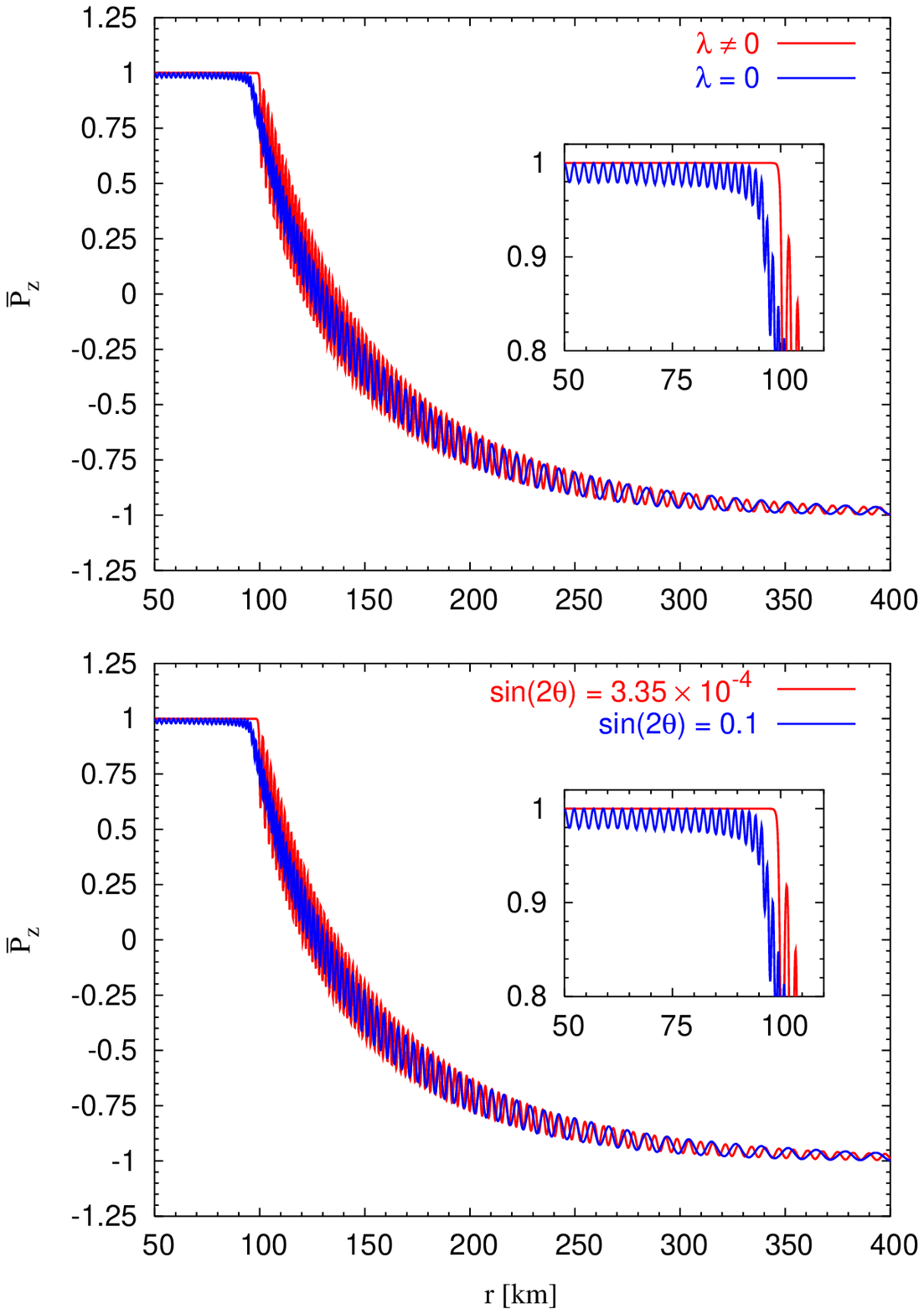}
\caption{Evolution for our standard inverted-hierarchy case for a
large vacuum mixing angle of $\sin2\theta=0.1$ (blue/dotted lines in
both panels), compared with different ways of suppressing the mixing
angle (red/solid). Top: ordinary matter effect according to the
profile at $t=2\,$s in Fig.~\ref{fig:profiles}. Bottom: small vacuum
mixing angle of
$\sin2\theta=3.35\times10^{-4}$.\label{fig:matterangle}}
\end{figure}

We observe that for the shown density profiles, the line $\omega$
intersects $\lambda(r)$ at a radius far exceeding the dense-neutrino
region that lies within the radius where the $\mu(r)$ profile
intersects $\omega$. In other words, the H-resonance is far outside
the region of interest except perhaps for very late times. Then, of
course, the neutrino luminosity will be much smaller, i.e., the
$\mu(r)$ curve would also shift downward and the dense-neutrino
region would be limited to smaller radii.

The true density profiles may be much lower, especially at late
times. This is even required for successful r-process
nucleosynthesis. In this scenario an MSW resonance may take place
within the dense-neutrino region, a case that was the focus of
previous numerical studies~\cite{Duan:2006an, Duan:2006jv}. However,
we will always assume that the H-resonance is at larger radii and
that neutrino-neutrino refraction and ordinary matter effects do not
interfere.

What is the impact of a large matter density in the region where
neutrino-neutrino effects are important? In the previous literature
it was recognized that a constant matter profile essentially reduces
the effective mixing angle so that matter should have the same
influence as a small vacuum mixing angle~\cite{Duan:2005cp,
Hannestad:2006nj}. We illustrate this point in
Fig.~\ref{fig:matterangle} with the evolution of $\bar P_z$ for our
usual case, but assuming now a large vacuum mixing angle $\sin
2\theta=0.1$. In the synchronization region one can now see
oscillations. We overlay this curve with $\bar P_z$, using the
matter profile of Fig.~\ref{fig:profiles} at $t=2\,$s. As expected,
matter has the effect of slightly delaying the onset of pair
transformations and of increasing the depth of the nutation
amplitude.

Actually, in the inverted hierarchy, the value of $\sin2\theta$ is
only crucial at the onset of the bipolar oscillations. Once the
overall polarization vector is tilted away from ${\bf B}$, the
initial ``misalignment'' with ${\bf B}$ no longer matters.
Therefore, what is crucial for the role of matter is only its
density around the region where synchronization ends. The in-medium
mixing angle at $r_{\rm synch}$ for the case shown in the top panel
of Fig.~\ref{fig:matterangle} is $\sin2\theta_{\rm
matter}=3.35\times10^{-4}$, assuming $\sin2\theta_{\rm vac}=0.1$.
Using this value of $\theta_{\rm matter}$ as a vacuum mixing angle
instead of matter yields the result shown in the bottom panel, again
overlaid with the original vacuum case of $\sin2\theta=0.1$.

We conclude that indeed we can ignore matter entirely if we account
for it schematically by a small vacuum mixing angle, at least in the
inverted hierarchy. Moreover, the onset of collective pair
transformations is only mildly changed by the choice of mixing
angle. Its main impact is that it controls the depth of the nutation
pattern. The exact matter profile is only important if it is so
shallow that it causes an MSW resonance in the dense-neutrino
region, a case that we do not investigate.

\subsection{Mixing angle}                         \label{sec:mixangle}

This discussion suggests that, at least for the inverted hierarchy,
the actual vacuum mixing angle does not strongly influence the issue
of multi-angle decoherence because this effect happens when the
global polarization vector is tilted far away from the ${\bf B}$
direction. On the other hand, we have already noted that the
quasi-coherent spiral structure that forms just beyond the
synchronization radius has more windings for a smaller mixing angle
so that the system is not identical.

\begin{figure}[ht]
\includegraphics[angle=0,width=0.95\columnwidth]{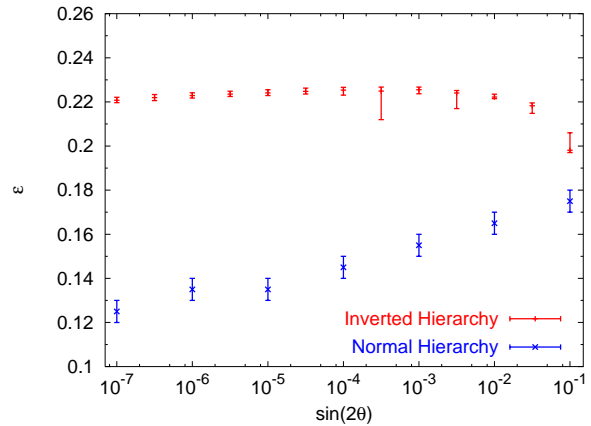}
\caption{Limiting $\epsilon$ for decoherence as a function of mixing
angle for our standard example and both
hierarchies.\label{fig:mixeps}}
\end{figure}

To clarify the role of the mixing angle we have used our standard
inverted-hierarchy case and have calculated the limiting asymmetry
$\epsilon$ for decoherence for a broad range of mixing angles. We show
the limiting contours in the plane of $\epsilon$ and $\sin 2\theta$ in
Fig.~\ref{fig:mixeps} for both hierarchies, above which
multiangle decoherence does not appear.

We emphasize that the limiting $\epsilon$ shown in
Fig.~\ref{fig:mixeps} has a different meaning for the two hierarchies.
As discussed earlier, in the inverted hierarchy, $\bar P$ shortens
somewhat even in the quasi single-angle regime. Therefore, as a formal
criterion for distinguishing the regions of coherence and decoherence
we use that the final $\bar P$ has shortened to less than 0.85. The
exact choice is irrelevant because the transition between the
quasi-coherent and decoherent regimes is steep as a function
of~$\epsilon$.

Conversely, in the normal hierarchy, $\bar P$ need not visibly shorten
at all as illustrated by the example in the lower left panel of
Fig.~\ref{fig:example2}. Therefore, we here demand that $\bar P$ does
not visibly shorten in such a picture. We construct the demarcation
line by decreasing $\epsilon$ in steps of 0.01 until the polarization
vector for the first time shortens visibly. Finding this point
requires a significant amount of manual iterations with a modified
inner radius and number of angular bins to make sure the result does
not depend on these numerical parameters. The error bars represent our
confidence range for the true critical value.

We conclude that for the inverted hierarchy, multi-angle decoherence
is virtually independent from the value of $\sin2\theta$, except
that for very large $\theta$ a slightly smaller asymmetry is enough
to suppress decoherence. Assuming the presence of ordinary matter,
such large mixing angles seem irrelevant, except perhaps at late
times. Either way, it is conservative to assume a small mixing angle
and we will use $\sin2\theta=10^{-3}$ as a default value.

For the normal hierarchy we find a strong dependence of the critical
$\epsilon$ on $\log_{10}(\sin2\theta)$. For a smaller mixing angle
it is easier to suppress decoherence. The normal hierarchy is very
different from the inverted one in that for a small mixing angle,
all polarization vectors stay closely aligned with the $z$-direction
unless multi-angle decoherence takes place. Therefore, it is
plausible that for a smaller mixing angle, decoherence effects are
delayed.

\subsection{Energy distribution}                    \label{sec:energy}

\begin{figure*}
\includegraphics[width=0.855\textwidth]{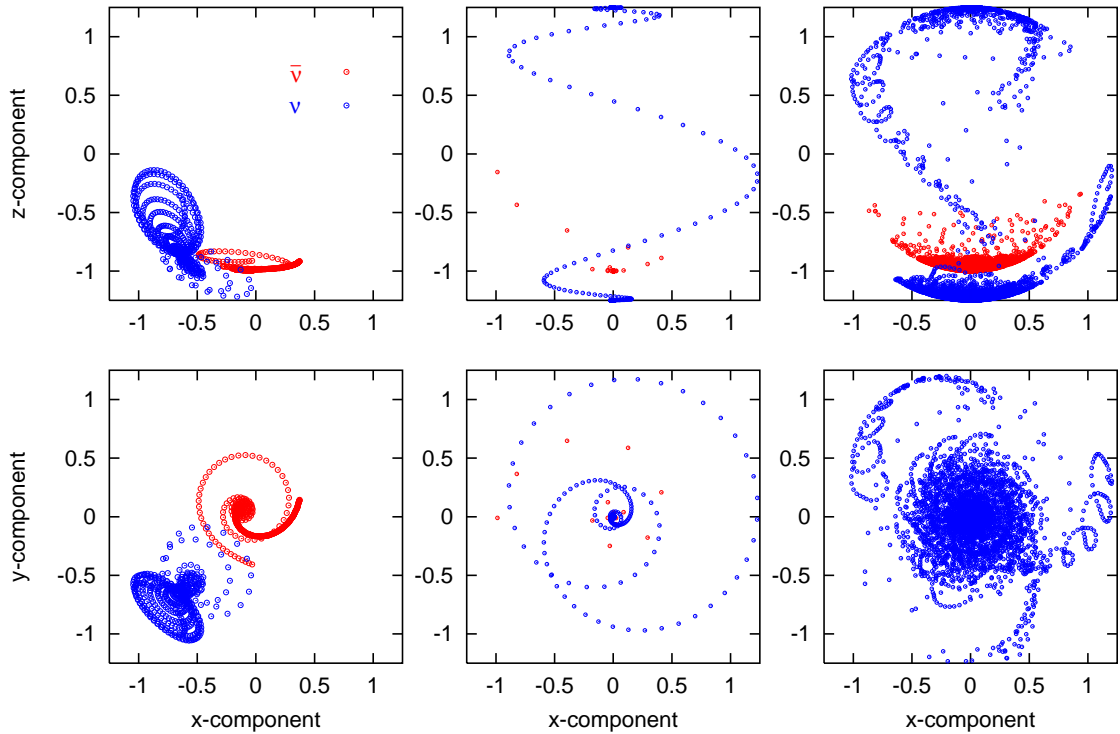}
\caption{Final state at a large radius of the polarization vectors
for our standard parameters in analogy to
Fig.~\ref{fig:footprints1}. The antineutrinos (red/light gray) are on the unit
sphere, whereas the neutrinos (blue/dark gray) live on a sphere of radius
$1+\epsilon=1.25$. Left: monochromatic multi-angle, the
antineutrinos being identical with the left column of
Fig.~\ref{fig:footprints1}. Middle: Box-like energy spectrum and
single angle. Right: Box-like energy spectrum and multi angle. In
the lower right panel we do not show the
antineutrinos.\label{fig:footprints3}}
\end{figure*}

The neutrinos emitted from a SN core naturally have a broad energy
distribution. In Ref.~\cite{Raffelt:2007yz} it was noted that the
energy distribution of neutrinos and antineutrinos is largely
irrelevant for the question of decoherence as long as the
oscillations exhibit self-maintained
coherence~\cite{Kostelecky:1995dt}. The multi-angle transition to
decoherence typically occurs within the dense-neutrino region where
the synchronization of energy modes remains strong. Therefore, we
expect that multi-angle decoherence is not significantly affected by
the neutrino spectrum.

In order to compare a monochromatic system with one that has a broad
energy distribution, the crucial quantity to keep fixed is not the
average energy, but the average oscillation frequency
$\langle\omega\rangle=\langle\Delta m^2/2E\rangle$. If we assume
that neutrinos and antineutrinos have equal distributions, it is
straightforward to adjust, for example, the temperature of a thermal
distribution such that $\langle\omega\rangle$ is identical to our
monochromatic standard case $\omega_0=0.3~{\rm km}^{-1}$.

If we assume different distributions for neutrinos and
antineutrinos, the equivalent $\omega_0$ is somewhat more subtle.
Consider first two different monochromatic spectra for neutrinos
with a fixed frequency $\omega_1$, and one for antineutrinos with a
different frequency $\omega_2$ (``bichromatic system''). Following
Ref.~\cite{Duan:2005cp} one can return to a monochromatic situation
by going into a reference frame that rotates around ${\bf B}$ with
such a frequency that in vacuum ${\bf P}$ and $\bar{\bf P}$ precess
around ${\bf B}$ with equal frequencies $\omega_0$, but in opposite
directions. The rotation frequency for the corotating frame is
$\omega_{\rm c}=(\omega_1-\omega_2)/2$. Therefore, our bichromatic
system behaves equivalently to a monochromatic one with
$\omega_0=(\omega_1+\omega_2)/2$. It is trivial to show in numerical
examples that the bichromatic system is indeed equivalent to a
monochromatic one with $\omega_0$ taken as the simple average of
$\omega_1$ and $\omega_2$.

If we have different distributions for neutrinos and antineutrinos,
we define the initial average frequencies by
\begin{equation}
 \langle\omega_\nu\rangle=
 \frac{\int_0^\infty\D\omega\,\omega\,P_\omega^z}
 {\int_0^\infty\D\omega\,P_\omega^z}\,,
\end{equation}
and analogous for $\langle\omega_{\bar\nu}\rangle$. The equivalent
monochromatic frequency is then $\omega_0=\frac{1}{2}\,
(\langle\omega_\nu\rangle+\langle\omega_{\bar\nu}\rangle)$. The
initial distribution $P_\omega^z$ can involve negative values if some
part of the spectrum initially consists of $\nu_x$ and not $\nu_e$.
Such spectral cross-overs occur, for example, if one assumes thermal
fluxes with equal luminosities but different temperatures.

We have studied several numerical examples of quasi single-angle
behavior and of multi-angle decoherence, taking different neutrino and
antineutrino energy spectra, such as flat or thermal and with equal or
different temperatures. We always found that the evolution of the
global polarization vectors is almost identical to the equivalent
monochromatic cases. We never observed that a broad energy spectrum
caused a significant deviation from the monochromatic behavior at
those radii that are relevant for decoherence.

Of course, a multi-energy system is qualitatively different from a
monochromatic one in that the final energy distribution shows a
``spectral split''~\cite{Duan:2006an, Duan:2006jv, Duan:2007mv,
Raffelt:2007cb, Mirizzi2007}. In a single-angle multi-energy system,
this means the $\bar\nu_e$ spectrum is completely transformed to the
$\bar\nu_x$ flavor, whereas only the high-energy part of the $\nu_e$
spectrum is transformed, the low-energy part remaining in (or rather
returning to) the original flavor. The energy $E_{\rm split}$ of
this sharp transition is fixed by lepton-number conservation in the
sense that the neutrino-neutrino interactions only catalyzes the
transformation of $\nu_e\bar\nu_e$ pairs. For various examples we
find results in full agreement with the previous
literature~\cite{Duan:2006an, Duan:2006jv, Duan:2007mv,
Raffelt:2007cb, Mirizzi2007}.

For sufficiently large asymmetries $\epsilon$ where the multi-angle
system evolves in the quasi single-angle mode, there is no
significant modification of the spectral split so that it is not
worthwhile to show any examples. In the decoherent case, the final
spectra naturally are very different, but we have not explored such
cases systematically because multi-angle decoherence does not seem
to be generic for realistic SN scenarios.

To illustrate the modifications caused by an energy spectrum in a
different way from the previous literature, we show in
Fig.~\ref{fig:footprints3} the side and top views of the location of
neutrino and antineutrino polarization vectors on the unit sphere in
analogy to Fig.~\ref{fig:footprints1} for our standard parameter
values. In the left column we show the same monochromatic multi-angle
case that we already showed in the left column of
Fig.~\ref{fig:footprints1}, with 500 modes. In addition we include the
neutrinos (blue/dark gray) that here live on a sphere of radius
$1+\epsilon=1.25$. The neutrinos form a spiral structure similar to
the one of the antineutrinos, but in the final state this structure
cannot move to the negative ${\bf B}$ directions because of lepton
number conservation.

In the middle column we show a single-angle example with the same
parameters, now using a box-like spectrum of oscillation frequencies
where initially $\bar P_{\omega}^z=(2\omega_0)^{-1}$ and
$P_{\omega}^z=(1+\epsilon)(2\omega_0)^{-1}$ for
$0\leq\omega\leq2\omega_0$ so that $\langle\omega_\nu\rangle
=\langle\omega_{\bar\nu}\rangle=\omega_0$ and it is equivalent to
the original monochromatic case. We now see that most of the
antineutrinos have moved to the negative ${\bf B}$ direction as
before, whereas the neutrinos populate both the positive and
negative ${\bf B}$ direction, representing the spectral split. The
lack of full adiabaticity prevents the split from being complete,
leaving some polarization vectors not fully aligned or anti-aligned
with ${\bf B}$. At large radii when the neutrino-neutrino
interactions have died out, these modes precess with their different
vacuum oscillation frequencies so that they are found on a spiral
locus extending from the ``south pole'' to the ``north pole'' that
gets wound up further at larger radii. Note that here we have used
1000 energy modes in order to obtain a visible population occupying
these non-adiabatic final states. Still, only very few red dots
(antineutrinos) are visible, the vast majority being at the south
pole. Likewise for the neutrinos (blue dots), the spiral is
populated only by a small fraction of the 1000 modes. In other
words, the evolution is nearly adiabatic.

Finally we combine a box-like energy spectrum and a multi-angle
distribution (right panels). The antineutrinos all cluster around
the negative ${\bf B}$ direction and fill the ``southern polar cap''
more or less uniformly because at late times modes with different
energies precess with different frequencies. The neutrinos populate
both the northern and southern polar caps, representing the spectral
split. At intermediate latitudes we find coherent spiral structures.
They correspond to modes with different angles but equal $\omega$ so
that even at late times they do not dissolve by differential
precession.

\subsection{Effective interaction strength} \label{sec:border}

\begin{figure}[ht]
\includegraphics[width=1.0\columnwidth]{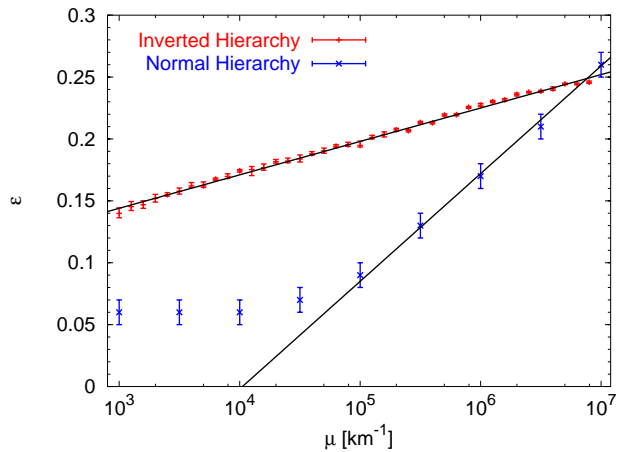}
\caption{Limiting $\epsilon$ for decoherence as a function of the
  effective neutrino-neutrino interaction strength $\mu$ for our
  standard parameters. The linear regressions are ``visual fits''
  represented by Eq.~(\ref{eq:epscontour}).\label{fig:regimes}}
\end{figure}

Besides the asymmetry $\epsilon$ itself, the most uncertain model
parameter is the effective neutrino-neutrino interaction strength
$\mu$ as defined in Eq.~(\ref{eq:mudefine}). In
Fig.~\ref{fig:regimes} we show the demarcation lines between
coherence and decoherence for both hierarchies in the
$\mu$-$\epsilon$-plane, keeping all other parameters at their
standard values. The contours are constructed as described in
Sec.~\ref{sec:mixangle}. The numerical contours are visually very
well approximated by linear regressions of the form
\begin{eqnarray}\label{eq:epscontour}
 \epsilon_{\rm IH}&\approx&0.225+0.027\,\log_{10}
 \left(\frac{\mu}{10^6~{\rm km}^{-1}}\right)\,,
 \nonumber\\*
 \epsilon_{\rm NH}&\approx&0.172+0.087\,\log_{10}
 \left(\frac{\mu}{10^6~{\rm km}^{-1}}\right)\,.
\end{eqnarray}
For the normal hierarchy, the linear regression would intersect
$\epsilon=0$ within the range of investigated $\mu$-values, but in
reality turns over and saturates around $\epsilon=0.06$.

\subsection{Vacuum oscillation frequency} \label{sec:omega}

The average vacuum oscillation frequency $\omega$ depends on the
atmospheric $\Delta m^2$ that is quite well constrained, and a
certain average of the neutrino energies. Our standard value is
$\omega=0.3~{\rm km}^{-1}$. If we increase this to $1~{\rm
km}^{-1}$, the $\epsilon$-$\mu$-contour in Fig.~\ref{fig:regimes} is
essentially parallel-shifted to larger $\epsilon$ by about 0.035
(inverted hierarchy). This range of $\omega$ probably brackets the
plausible possibilities so that the uncertainty of $\omega$ does
not strongly influence the practical demarcation between the
regimes.

\begin{figure}[ht]
\includegraphics[width=1.0\columnwidth]{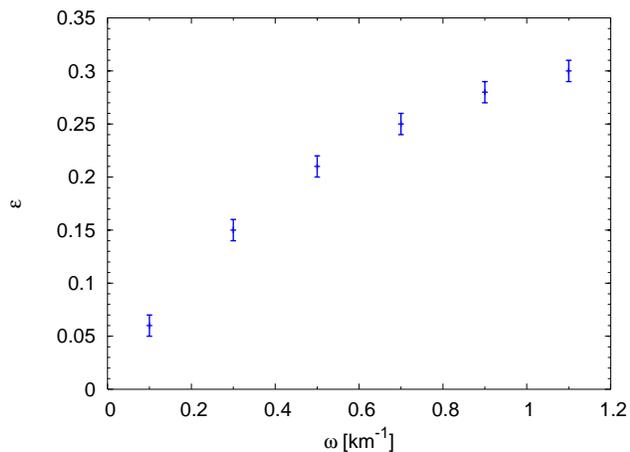}
\caption{Limiting $\epsilon$ for decoherence as a function of
  the vacuum oscillation frequency $\omega$ for our standard
  parameters and the normal hierarchy.\label{fig:nhomega}}
\end{figure}

The normal hierarchy is more sensitive to $\omega$.  In
Fig.~\ref{fig:nhomega} we show a contour for the coherence regime in
the $\epsilon$-$\omega$ plane, assuming otherwise our standard
parameter values. Changing $\omega$ from $0.3$ to $1~{\rm km}^{-1}$
increases the critical $\epsilon$ by almost~0.15.

\section{Conclusions}
\label{sec:conclusions}

The nonlinear neutrino transformations that occur in the
dense-neutrino region of a SN show numerous novel features. It was
noted that multi-angle effects play an important role in that the
neutrino-neutrino interaction depends on the relative angles of the
various trajectories~\cite{Duan:2006an, Duan:2006jv}. At the same
time it was numerically observed that for a typical example the
behavior was unexpectedly quite similar to the single-angle
case~\cite{Duan:2006an, Duan:2006jv}. On the other hand it was
analytically shown that a gas of equal densities of neutrinos and
anti-neutrinos has a pronounced angular instability and kinematical
decoherence between different angular modes in flavor space is fast,
representing a stable fixed point of the
system~\cite{Raffelt:2007yz}.

We have here not attempted to develop further analytical insights, but
have taken a practical approach and explored numerically the range of
parameters where different forms of behavior dominate in a realistic
SN scenario.

To this end we have first clarified that ``multi-angle effects''
mean one of two clearly separated forms of behavior. The flavor
content of the system can evolve in a quasi single-angle form. On
the level of the polarization vectors this means that they fill only
a restricted volume of the available phase space and maintain a
coherent structure. On the other hand, nearly complete flavor
equilibrium can arise where the available phase space is more or
less uniformly filled.

For realistic assumptions about supernova and neutrino parameters,
the switch between these modes of evolution is set by the degree of
asymmetry between the neutrino and antineutrino fluxes. While this
asymmetry is caused by the deleptonization flux, the crucial
parameter $\epsilon$ is the asymmetry between $F(\nu_e)-F(\nu_x)$
and the corresponding antineutrino quantity as defined in
Eq.~(\ref{eq:epsdefine}) because for flavor oscillations the part of
the density matrix that is proportional to $F(\nu_e)+F(\nu_x)$ drops
out. While in a realistic SN on average $F(\nu_e)$ is about 15\%
larger than $F(\bar\nu_e)$, the asymmetry parameter as defined in
Eq.~(\ref{eq:epsdefine}) is typically much larger. 

The critical value of $\epsilon$ that is enough to suppress
decoherence depends on the type of neutrino mass hierarchy, the
average energies, luminosities, and on the mixing angle. We have
found that for $\epsilon\agt0.3$, decoherence is suppressed for the
entire range of plausible parameters, but a value smaller than 0.1
may be enough, depending on the combination of other parameters.

We conclude that the quasi single-angle behavior may well be typical
for realistic SN conditions, i.e., that the deleptonization flux is
enough to suppress multi-angle decoherence. To substantiate this
conclusion one should analyze the output of numerical simulations in
terms of our model parameters. Besides the flavor-dependent
luminosities and average energies, one needs the angular
distribution of the neutrino radiation field at some radius where
collisions are no longer important.

If our conclusion holds up in the light of realistic SN simulations,
a practical understanding of the effect of self-induced neutrino
flavor transformations quickly comes into reach. In the normal mass
hierarchy, nothing new would happen on a macroscopic scale. In the
inverted hierarchy, the final effect would be a conversion of
$\nu_e\bar\nu_e$ pairs and a split in the $\nu_e$ spectrum. These
phenomena are only mildly affected by multi-angle effects as long as
we are in the quasi single-angle regime.

If at late times the matter density profile contracts enough that an
MSW effect occurs in the dense-neutrino region, the situation becomes
more complicated as the neutrino-neutrino and ordinary matter effects
interfere and produce a richer structure of spectral
modifications~\cite{Duan:2006an, Duan:2006jv}.  Even then, numerical
simulations are much simpler if multi-angle decoherence is suppressed.

It is not obvious how $\epsilon$ evolves at late times. The
deleptonization of the core is probably faster than the cooling so
that one may think that $\epsilon$ becomes smaller. On the other hand,
the $\bar\nu_e$ can essentially only interact via neutral current
reactions and their flux and energy distribution should, therefore,
become very similar to the ones of $\nu_x$ and $\bar\nu_x$. Therefore,
it is not obvious if at late times the initial flux difference
$F_{\nu_e}-F_{\bar\nu_e}$ or $F_{\bar\nu_e}-F_{\bar\nu_x}$ decreases
more quickly. We also note that there can be a cross-over in the sense
that at late times the flux hierarchy can become
$F_{\nu_x}=F_{\bar\nu_x} >F_{\nu_e}>F_{\bar\nu_e}$ as in
Ref.~\cite{Raffelt:2003en}, meaning that we would have a pair excess
flux of $\nu_x\bar\nu_x$ instead of a $\nu_e\bar\nu_e$ excess.

We have refrained from an interpretation of our numerical findings
because we do not have developed a theory of kinematical decoherence
for a system that is asymmetric between neutrinos and antineutrinos
and where the effective interaction strength varies as a function of
time (or here of radius). The absence of multi-angle decoherence
seems to follow from a lack of time for it to develop. One can
interpret our results such that a more adiabatic decrease of the
neutrino-neutrino interaction strength requires a larger asymmetry
to suppress decoherence. The different length scales of the problem
seem to conspire such that the evolution is adiabatic in that sharp
spectral splits develop, but not so adiabatic that kinematical
decoherence would be typical. An analytic understanding of this
conspiracy remains to be found.

Our results suggest that signatures of collective flavor
transformations are not erased by multi-angle decoherence and will
survive to the surface, modulated by the usual MSW flavor
conversions~\cite{Dighe:1999bi}. The survival of observable signatures
then also depends on the density fluctuations of the ordinary medium
that can be a source of kinematical flavor
decoherence~\cite{Fogli:2006xy,Friedland:2006ta}.

All authors in this field have relied on the simplifying assumption of
either homogeneity or exact spherical symmetry to make the equations
numerically tractable. The neutrino emission from a real SN is
influenced by density and temperature fluctuations of the medium in
the region where neutrinos decouple. Likewise, the neutrino fluxes
emitted from the accretion tori of coalescing neutron stars, the
likely engines of short gamma ray bursts, have fewer symmetries than
assumed here. It remains to be investigated if systems with more
general geometries behave qualitatively similar to the spherically
symmetric case or if deviations from spherical symmetry can provide a
new source of kinematical decoherence.


\begin{acknowledgments}
This work was partly supported by the Deutsche
Forschungsgemeinschaft under the grant TR-27 ``Neutrinos and
Beyond'', by The Cluster of Excellence for Fundamental Physics
``Origin and Structure of the Universe'' (Garching and Munich), by
the European Union under the ILIAS project (contract No.\
RII3-CT-2004-506222) and an RT Network (contract No.\
MRTN-CT-2004-503369), and by the Spanish grants FPA2005-01269 (MEC)
and ACOMP07-270 (Generalitat Valenciana). A. E. has been supported by
a FPU grant from the Spanish Government. SP and RT were supported
by MEC contracts ({\em Ram\'{o}n y Cajal} and {\em Juan de la
Cierva}, respectively).
\end{acknowledgments}

\appendix
\section{Equations of motion}
\label{app:eom}

\subsection{Temporal evolution}

A homogeneous ensemble of unmixed neutrinos is represented by the
occupation numbers $f_{\bf p}=\langle a^\dagger_{\bf p} a_{\bf
p}\rangle$ for each momentum mode ${\bf p}$, where $a^\dagger_{\bf
p}$ and $a_{\bf p}$ are the relevant creation and annihilation
operators and $\langle\ldots\rangle$ is the expectation value. A
corresponding expression can be defined for the antineutrinos,
${\bar f}_{\bf p}=\langle {\bar a}^\dagger_{\bf p} {\bar a}_{\bf
p}\rangle$, where overbarred quantities always refer to
antiparticles. In a multiflavor system of mixed neutrinos, the
occupation numbers are generalised to density matrices in flavor
space~\cite{Dolgov:1980cq, Sigl:1992fn, mckellar&thomson}
\begin{equation}\label{eq:densitymatrixdefinition}
 (\varrho_{\bf p})_{ij}=
 \langle a^\dagger_{i} a_{j}\rangle_{\bf p}
 \hbox{\quad and\quad}
 (\bar \varrho_{\bf p})_{ij}=
 \langle \bar a^\dagger_{j}\,\bar a_{i}\rangle_{\bf p}\,.
\end{equation}
The reversed order of the flavor indices $i$ and $j$ in the
right-hand side for antineutrinos assures that $\varrho_{\bf p}$ and
$\bar\varrho_{\bf p}$ transform identically under a flavor
transformation.

Flavor oscillations of an ensemble of neutrinos and antineutrinos
are described by~\cite{Dolgov:1980cq, Sigl:1992fn, mckellar&thomson}
\begin{equation}\label{eq:eom1}
 \I\partial_t\varrho_{\bf p}=[{\sf H}_{\bf p},\varrho_{\bf p}]
 \hbox{\quad and\quad}
 \I\partial_t\bar\varrho_{\bf p}=
 [\bar {\sf H}_{\bf p},\bar\varrho_{\bf p}]\,,
\end{equation}
where $[{\cdot},{\cdot}]$ is a commutator. The ``Hamiltonian'' for
each mode is
\begin{equation}\label{eq:ham1}
 {\sf H}_{\bf p}=\Omega_{\bf p}
 +\lambda {\sf L}+\sqrt{2}\,G_{\rm F}
 \int\!\frac{\D^3{\bf q}}{(2\pi)^3}
 \left(\varrho_{\bf q}-\bar\varrho_{\bf q}\right)
 (1-{\bf v}_{\bf q}\cdot{\bf v}_{\bf p})\,,
\end{equation}
where $G_{\rm F}$ is the Fermi constant. The matrix of vacuum
oscillation frequencies for relativistic neutrinos is in the mass
basis $\Omega_{\bf p}={\rm diag}(m_1^2,m_2^2,m_3^2)/2p$ with
$p=|{\bf p}|$. The matter effect is represented by
$\lambda=\sqrt{2}\,G_{\rm F}(n_{e^-}-n_{e^+})$ and ${\sf L}={\rm
diag}(1,0,0)$, given here in the weak interaction basis. We ignore
the possible presence of other charged-lepton flavors. The
Hamiltonian for antineutrinos $\bar {\sf H}_{\bf p}$ is the same
with $\Omega_{\bf p}\to-\Omega_{\bf p}$, i.e., in vacuum
antineutrinos oscillate ``the other way round.''

The factor $(1-{\bf v}_{\bf q}\cdot{\bf v}_{\bf
p})=(1-\cos\theta_{\bf pq})$ represents the current-current nature
of the weak interaction where ${\bf v}_{\bf p}={\bf p}/p$ is the
velocity. The angular term averages to zero if the gas is isotropic.
We ignore a possible net flux of charged leptons lest the ordinary
matter effect also involves an angular factor.

If the system is axially symmetric relative to some direction, the
angular factor simplifies after an azimuthal integration
to~\cite{Duan:2006an, Raffelt:2007yz}
\begin{equation}\label{eq:axial}
 (1-{\bf v}_{\bf q}\cdot{\bf v}_{\bf p})
 \to(1-{v}_{\bf q}{v}_{\bf p})\,,
\end{equation}
where the velocities are along the symmetry axis.

\subsection{Spatial evolution in spherical symmetry}

Instead of a homogeneous system that evolves in time we consider a
stationary system that evolves in space. The occupation numbers
become Wigner functions, which depend both on spatial coordinates
and on momenta, but there is no conceptual problem as long as we
consider spatial variations that are slow on the scale of the
inverse neutrino momenta.

Since multi-angle effects are at the focus of our problem, we cannot
reduce the equations to plane waves moving in the same direction.
Motivated by the SN application, however, we can take advantage of
global spherical symmetry, implying that the ensemble is represented
by matrices that depend on a radial coordinate $r$, the zenith angle
relative to the radial direction, and the energy $E$ which in the
relativistic limit is identical with $p=|{\bf p}|$.

We ignore gravitational deflection near the SN core and assume that
neutrinos move on straight lines after being launched at a radius
$R$ that we call the neutrino sphere. Consider a neutrino that was
launched at an angle $\vartheta_R$ relative to the radial direction.
Its radial velocity is
\begin{equation}
v_R=\cos\vartheta_R\,.
\end{equation}
At $r>R$ the trajectory's angle relative to the radial
direction is implied by simple geometry to be~\cite{Duan:2006an} (see
e.g. their Fig. 1)
\begin{equation}
R\sin\vartheta_R=r\sin\vartheta_r\,.
\end{equation}
Therefore, the radial velocity at $r$ is
\begin{equation}\label{eq:costheta}
v_{u,r}=\cos\vartheta_r=\sqrt{1-\frac{R^2}{r^2}\,u}
\end{equation}
where we have introduced
\begin{equation}\label{eq:udef}
u=1-v^2_R=\sin^2\vartheta_R\,.
\end{equation}
It is convenient to label the angular modes with $u$. The physical
zenith angles change with distance so that the equations would be
more complicated.

The density matrices $\varrho_{p,u,r}$ are not especially useful to
describe a spherically symmetric system because they vary with $r$
even in the absence of oscillations. (Note that we often write the
dependence of a quantity on a variable as an subscript.) A quantity
that is conserved in the absence of oscillations is the total flux
matrix
\begin{equation}
  {\sf J}_r=\frac{r^2}{R^2}
  \int\frac{\D^3{\bf p}}{(2\pi)^3}\,
  \varrho_{{\bf p},r}\,v_{{\bf p},r}\,.
\end{equation}
To express the integral in co-moving variables we observe that
$\D^3{\bf p}$ in spherical coordinates is $p^2\D
p\,\D\varphi\,\D\cos\vartheta_r$ and that Eq.~(\ref{eq:costheta})
implies
\begin{equation}\label{eq:transformation}
 \left|\frac{\D\cos\vartheta_r}{\D u}\right|=
 \frac{1}{2v_{u,r}}\,\frac{R^2}{r^2}\,.
\end{equation}
Therefore, we finally define the differential flux matrices
\begin{equation}\label{eq:Judef}
 {\sf J}_{p,u,r}=
 \frac{p^2\varrho_{p,u,r}}{2\,(2\pi)^2}\,,
\end{equation}
where we have used $\int\D\varphi=2\pi$ for axial symmetry. The
normalization is
\begin{equation}
{\sf J}_r=\int_0^1\D u\int_0^\infty\D p\; {\sf J}_{p,u,r}\,.
\end{equation}
In the absence of oscillations the total and differential fluxes are
conserved, $\partial_r {\sf J}_{r}=0$ and $\partial_r {\sf
J}_{p,u,r}=0$.

To include oscillations, we note that the radial velocity along a
neutrino trajectory is $v_{u,r}=\D r_{u}/\D t=\cos\vartheta_{u,r}$.
Therefore, if we wish to express the temporal evolution of the
neutrino density matrix along its trajectory in terms of an
evolution expressed in terms of the radial coordinate~$r$, we
substitute $\partial_t\to v_{u,r}\partial_r$ in Eq.~(\ref{eq:eom1})
so that
\begin{equation}
 {\rm i}\partial_r{\sf J}_{p,u,r}=
 \frac{[{\sf H}_{p,u,r},{\sf J}_{p,u,r}]}{v_{u,r}}
 \,,
\end{equation}
and analogous for antineutrinos. In other words, we project the
evolution along a given trajectory to an evolution along the radial
direction. For vacuum oscillations this has the effect of
``compressing'' the oscillation pattern for non-radial modes, i.e.,
even for monochromatic neutrinos, the effective vacuum oscillation
frequency depends on both $r$ and $u$.

The vacuum-oscillation and ordinary-matter contributions to ${\sf
H}_{p,u,r}$ were given in Eq.~(\ref{eq:ham1}), whereas the self-term
must be made explicit. To this end we introduce the matrix of number
densities
\begin{equation}
{\sf N}_{p,u,r}=v_{u,r}^{-1}\,{\sf J}_{p,u,r}
\end{equation}
and its integral as
\begin{equation}
 {\sf N}_{r}=\int_0^1 \D u\int_0^\infty\D p\;{\sf N}_{p,u,r}
 =\int_0^\infty\D u\int_0^\infty\D p\;
 \frac{{\sf J}_{p,u,r}}{v_{u,r}}\,.
\end{equation}
Collecting all terms and taking advantage of Eq.~(\ref{eq:axial})
for axial symmetry, we find
\begin{widetext}
\begin{eqnarray}\label{eq:eom0}
 \I\partial_r{\sf J}_{p,u,r}&=&+\bigl[\Omega_p,{\sf N}_{p,u,r}\bigr]
 +\lambda_r\bigl[{\sf L},{\sf N}_{p,u,r}\bigr]
 +\sqrt{2}G_{\rm F}\,\frac{R^2}{r^2}\,
 \Bigl(\bigl[{\sf N}_r-\bar{\sf N}_r,{\sf N}_{p,u,r}\bigr]
 -\bigl[{\sf J}_r-\bar{\sf J}_r,{\sf J}_{p,u,r}\bigr]\Bigr)\,,
 \nonumber\\*
 \I\partial_r\bar{\sf J}_{p,u,r}&=&
 -\bigl[\Omega_p,\bar{\sf N}_{p,u,r}\bigr]
 +\lambda_r\bigl[{\sf L},\bar{\sf N}_{p,u,r}\bigr]
 +\sqrt{2}G_{\rm F}\,\frac{R^2}{r^2}\,
 \Bigl(\bigl[{\sf N}_r-\bar{\sf N}_r,\bar{\sf N}_{p,u,r}\bigr]
 -\bigl[{\sf J}_r-\bar{\sf J}_r,\bar{\sf J}_{p,u,r}\bigr]\Bigr)\,,
\end{eqnarray}
\end{widetext}
where the electron density's radial variation is included
in~$\lambda_r$.

\subsection{Angular emission characteristics}

In a numerical simulation we need to specify the fluxes at the
neutrino sphere $r=R$. For our usual multi-angle simulations we
assume that the neutrino radiation field is ``half isotropic''
directly above the neutrino sphere, i.e., that all outward-moving
angular modes are equally occupied as behooves a thermal radiation
field. Therefore, the occupation numbers are distributed as $\D
n/\D\cos\vartheta_R={\rm const.}$, implying that the radial fluxes
are distributed as $\D j/\D\cos\vartheta_R=v_R\D
n/\D\cos\vartheta_R\propto\cos\vartheta_R$ because
$v_R=\cos\vartheta_R$. Expressed in the angular variable $u$ this
implies $\D j/\D u={\rm const.}$ because of
Eq.~(\ref{eq:transformation}). In other words, a blackbody radiation
field at the neutrino sphere implies that
\begin{equation}
{\sf J}_u={\rm const.}
\end{equation}
in the interval $0\leq u\leq1$.

To avoid multi-angle effects one may sometimes wish to use a single
angular bin. To represent a uniform ${\sf J}_u$ distribution, the
natural choice is $u=1/2$, corresponding to a launch angle
$\vartheta_R=45^\circ$. Our numerical single-angle examples always
correspond to this choice in an otherwise unchanged numerical code.

In this case the radial velocity of all neutrinos as a function of
radius is
\begin{equation}
v_r=\sqrt{1-\frac{R^2}{2r^2}}\,.
\end{equation}
For a monochromatic spectrum, the remaining flavor matrices are
simply the total ${\sf J}_r$ (corresponding to the single $u=1/2$)
and ${\sf N}_r={\sf J}_r/v_r$. Ignoring the trivial ordinary matter
term, the equations of motion are
\begin{equation}
 \I\partial_r{\sf J}_{r}=
 \frac{\bigl[\Omega,{\sf J}_{r}\bigr]}{v_r}
 +\sqrt{2}G_{\rm F}\,\frac{R^2}{r^2}\,
 \left(\frac{1}{v_r^2}-1\right)
 \bigl[{\sf J}_r-\bar{\sf J}_r,{\sf J}_{r}\bigr]
\end{equation}
and analogous for the antineutrinos. The coefficient of the
neutrino-neutrino term is explicitly
\begin{equation}\label{eq:muvariation}
 \sqrt{2}G_{\rm F}\,\frac{R^4}{r^4}\,\frac{1}{2-R^2/r^2}\,.
\end{equation}
At the neutrino sphere it is equal to $\sqrt{2}G_{\rm F}$, whereas
at large distances it is $(\sqrt{2}G_{\rm F}/2)\,R^4/r^4$. As
observed in the previous literature, the neutrino-neutrino term dies
out at large distances as $r^{-4}$.

One can define a ``single-angle case'' somewhat differently.
Assuming all angular modes evolve coherently, we can integrate the
equations of motion over $\int_0^1 \D u$ and study the evolution of
the quantities $J_{p,r}=\int_0^1\D u J_{p,u,r}$. To write the
equations in a compact form we introduce the notation
\begin{equation}
 \frac{1}{v_r^*}\equiv\frac{1}{{\sf J}_r}
 \int_0^\infty\!\D p\int_0^1\D u\;\frac{{\sf J}_{p,u,r}}{v_{u,r}}\,.
\end{equation}
The full equation of motion Eq.~(\ref{eq:eom0}) for neutrinos
becomes
\begin{eqnarray}\label{eq:eom0a}
 \I\partial_r{\sf J}_{p,r}&=&
 \frac{\bigl[\Omega_p,{\sf J}_{p,r}\bigr]}{v_r^*}
 +\lambda_r\,\frac{\bigl[{\sf L},{\sf J}_{p,r}\bigr]}{v_r^*}
 \nonumber\\
 &&{}+\sqrt{2}G_{\rm F}\,\frac{R^2}{r^2}\,
 \left(\frac{1}{(v_r^*)^2}-1\right)
 \bigl[{\sf J}_r-\bar{\sf J}_r,{\sf J}_{p,r}\bigr]
 \nonumber\\
\end{eqnarray}
and analogous for antineutrinos with $\Omega_p\to-\Omega_p$.

At large distances we have $1/v_r^*=1+\frac{1}{2}(R/r)^2\langle
u\rangle$ where $\langle u\rangle$ is the average of $u$ at
emission. For the vacuum and matter terms, we only need the leading
terms so that we recover the familiar plane-wave form of the
equations of motion. The coefficient of the neutrino-neutrino term,
on the other hand, becomes
\begin{equation}
\sqrt{2}G_{\rm F}\,\frac{R^4}{r^4}\,\langle u\rangle\,.
\end{equation}
Both for half-isotropic emission and for our single-angle case we
have $\langle u\rangle=\frac{1}{2}$, in agreement with
Eq.~(\ref{eq:muvariation}).


\end{document}